\documentstyle[aps,prl,times]{revtex}

\input epsf
\epsfverbosetrue

\begin{document} 
\draft
\sloppy
\twocolumn[
\hsize\textwidth\columnwidth\hsize\csname
@twocolumnfalse\endcsname

%%%%%%%%%%%%%%%%%%%%%%%%%%%%%%%%%%%%%%%%%%%%%%%%%%%%%%%%%%%%%%%%%%%%%%%%

\title{Shot Noise by Quantum Scattering in Chaotic Cavities\\}
\author{S.~Oberholzer, E.~V.~Sukhorukov, C.~Strunk, and C. Sch\"onenberger}
\address{Institut f\"ur Physik, Universit\"at Basel\\
Klingelbergstr.~82, CH-4056 Basel, Switzerland}

\author{T.~Heinzel and K.~Ensslin}
\address{Solid State Physics Laboratory, ETH-Z\"urich\\
CH-8093 Z\"urich, Switzerland}

\author{M.~Holland}
\address{Department of Electronics, University of Glasgow\\
Glasgow G12 8QQ, United Kingdom}

\date{\today}

\maketitle

\begin{abstract}
We have experimentally studied shot noise of chaotic 
cavities defined by two quantum point contacts in 
series. The cavity noise is determined as $1/4\cdot 
2e|I|$ in agreement with theory and can be 
well distinguished from other contributions 
to noise generated at the contacts. Subsequently,  
we have found that cavity noise decreases if one of the contacts is 
further opened and reaches nearly zero for a highly asymmetric cavity. 
\end{abstract}
]

\begin{figure}[h]
\centering
\epsfxsize=76 mm
\epsfbox{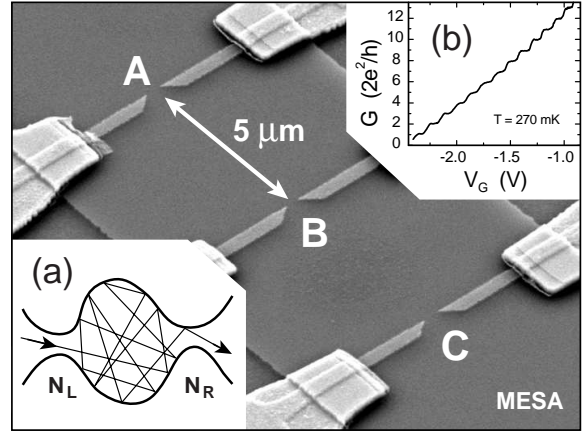}
\vspace{1.5mm}
\caption{SEM-picture of a Hall bar with three QPCs in series used to 
define chaotic cavities of different size. (a) The ratio of the number of modes
$\eta\equiv N_{L}/N_{R}=G_{L}/G_{R}$ can be adjusted by varying the 
openings of the left and right contact, independently. 
(b) QPC conductance vs. gate voltage of one of the contacts.}
\label{sample}
\end{figure}

The non-equilibrium time dependent fluctuations of the electrical 
current, known as shot noise, 
are caused by the randomness of charge transfer
in units of $e$ \cite{review}.
If the electron transfer can be described by
a Poissonian process, 
the spectral density $S$ 
of the current fluctuations 
is $S_{Poisson}=2e|I|$. 
Correlations imposed by fermionic statistics 
as well as Coulomb interaction may change 
shot noise from $S_{Poisson}$.
This is expressed by the Fano factor
$F$ defined as $F\equiv S/S_{Poisson}$.
A quantum wire with an intermediate
barrier with energy-independent 
transmission probability 
$\Gamma$, $\Gamma=1/2$ for example, has a Fano factor
of $1-\Gamma=1/2$ \cite{Lesovik1989,Buttiker1990}.
This suppression is due to binominal instead 
of Poissonian statistics.
Here, we explore what happens if the barrier is
replaced by a chaotic cavity (Fig.~\ref{sample}, inset (a)). 
For a symmetric and open cavity, which
is a cavity connected to ideal reservoirs via two
identical {\em noiseless} (barrier-free) quantum wires, the
mean transmission probability is $1/2$, too. 
But surprisingly, the Fano factor is predicted to be only $1/4$ 
\cite{Jalabert94,BSPRL2000}.
The $1/4$ Fano factor, valid for open and symmetric 
chaotic cavities, was first derived by
random matrix theory (RMT), which is based on phase-coherent 
quantum mechanical transport \cite{Jalabert94}.
Recently, a semiclassical analysis  
using the ``principle of minimal correlations'' arrived 
at the same result \cite{BSPRL2000}. 

Similar to metallic diffusive wires, 
where $F=1/3$ \cite{ButtikerBeenakker92,Schoelkopf97,onethird}, 
the Fano factor $1/4$ 
for  chaotic cavity is universal in the 
sense that it is insensitive to 
microscopic properties 
\cite{Jalabert94,BSPRL2000,Nazarov94,SukhorukovPRL}.
Nevertheless, there is an important difference 
between these two systems concerning the 
origin of resistance and noise. 
In a diffusive conductor resistance {\em and}
shot noise are both generated locally
at scattering centers, which are homogeneously distributed
along the wire.
In an open chaotic cavity resistance and shot noise
are generated differently. The
resistance is due to the 
fundamental
quantum resistance of the contacts. 
Although the source of resistance, the open contacts
do not contribute to noise because electrons are 
transmitted with unit probability.
Shot noise arises \emph{inside} the cavity 
due to quantum
\noindent mechanical diffraction 
which splits the electron wave 
packet into two partial waves leaving the two exits.
 In the semiclassical approach cavity noise is determined
by the average fluctutations of the state occupancy
inside the cavity given, at $T=0$, by \cite{BSPRL2000}
\begin{equation}
	S=2G\int dE f_{C}(1-f_{C}).
	\label{eq:fff}
\end{equation}
Here, $f_{C}(E)$ denotes the distribution function inside 
the cavity, which is homogeneous and isotropic. 
The total conductance
$G=G_{0}(N_{L}N_{R})/(N_{L}+N_{R})$ with $G_{0}=2e^2/h$ 
is equal to the series conductance of the left and right 
contact with $N_{L}$ ($N_{R}$) open channels (i.e. $\Gamma_{1\ldots N_{L,R}}=1$, 
$\Gamma_{>N_{L,R}}=0$).
For non-interacting electrons the distribution 
function in the cavity $f_{C}$  just equals 
the weighted average of the distribution functions 
$f_{L}$ and $f_{R}$ 
in the left and right reservoirs.
In the symmetric case 
$N_{L}=N_{R}$, i.e. $f_{C}=\frac{1}{2}(f_{L}+f_{R})$, and
Eq.~(\ref{eq:fff}) yields a Fano factor of $1/4$.
For very asymmetric contacts ($N_{L}\gg N_{R}$) shot noise 
approaches
zero, since the system can then be regarded as a single
contact with $N_{R}$ open and therefore noiseless channels.
The general Fano factor $F\equiv S/2e|I|$ for cavity noise is
\begin{equation}
	F(\eta) = \frac{N_{L}N_{R}}{(N_{L}+N_{R})^2} = \frac{\eta}{(1+\eta)^2},
	\label{eq:cavitynoisecold}
\end{equation}
where we introduce the parameter $\eta\equiv  N_{L}/N_{R}$ 
measuring the symmetry of the cavity.

Experimentally, we have realized chaotic cavities
by two quantum 
point contacts (QPC) in series.
These are electrostatically defined in a two dimensional electron gas (2 DEG) by 
metallic split gates on top (see Fig.~\ref{sample}) \cite{Wees1988}.
The opening of the contacts can be individually tuned by varying the applied gate 
voltages independently. 
The \mbox{2 DEG} forms 80 nm below the surface at the interface of a 
standard GaAs/Al$_{0.3}$Ga$_{0.7}$As-heterojunction.
Magnetoresistance measurements yield a carrier density of 
2.7$\times$10$^{15}$ m$^{-2}$, corresponding to a Fermi energy of 
\mbox{$\simeq$106 K} and a mobility of \mbox{83 Vs/m$^{-2}$} 
resulting in a mean 
free path of $\simeq$7 $\mu$m comparable to the size of the cavity. 
Three QPCs in series as shown in Fig.~\ref{sample} enable to define 
two cavities of different size: either the outer gates A and C with the middle 
gate B kept 
completely open can be used to define a relatively
large cavity  of $\simeq$11$\times$8 $\mu$m, or 2 of the 
inner gates (A,B or B,C) creating a smaller cavity of \mbox{$\simeq$5$\times$8 $\mu$m.}
The conductance of the QPCs is quantized according to the Landauer formula $G=G_{0} 
\sum_{n}\Gamma_{n}$ \cite{Landauer1957} (\mbox{inset (b)} of Fig.~\ref{sample})).
An \emph{open} cavity is defined when both QPCs 
are adjusted to a conductance plateau, 
where $N$ modes are fully transmitted ($\Gamma=1$) and the others are totally reflected 
($\Gamma=0$). 
The two-terminal conductance $G$ is experimentally found to correspond to the series 
conductance of the two contacts $G_{L}G_{R}/(G_{L}+G_{R})$ with an 
accuracy of less than $1\%$ \cite{Jalabert94,BSPRL2000}.  
Therefore, direct transmission of electrons from the left to the right 
contact can be excluded, as well as quantum corrections 
\cite{Marcus98,comment}.

Two independent low-noise amplifiers (EG\&G 5184) operating at room temperature
are used to detect the voltage fluctuations across the cavity.
A spectrum analyzer (HP 89410A) calculates the cross-correlation spectrum 
of the two amplified signals. This technique allows to reduce
uncorrelated noise contributions which do not originate from the sample itself.
Experimental details can be found  in \cite{onethird,Kumar96}.
Furthermore, the whole setup is filtered against  RF-interference 
 at low temperatures 
by a shielded sample-box and lossy  microcoaxes to minimize 
heating by radiation. 
Voltage noise is typically measured at frequencies around 6 kHz
where the noise is frequency independent (white) 
up to the maximum bias 
current $\le$50 nA used in the experiment.
The sensitivity for voltage noise measurements 
is of the order 5$\cdot$10$^{-21}$ V$^2$s.
The measured noise is calibrated against
equilibrium Nyquist noise at different bath-temperatures.
From the Nyquist-relation $S_{V}=4k_{B}RT$ the voltage 
gain as well as the offset in the voltage 
noise $S_{I}^{off}R^2$ caused by the finite current noise $S_{I}^{off}$ of the 
amplifiers can be determined with high accuracy. Although shot noise is a 
\begin{figure}[h]
\centering
\epsfxsize=86 mm
\epsfbox{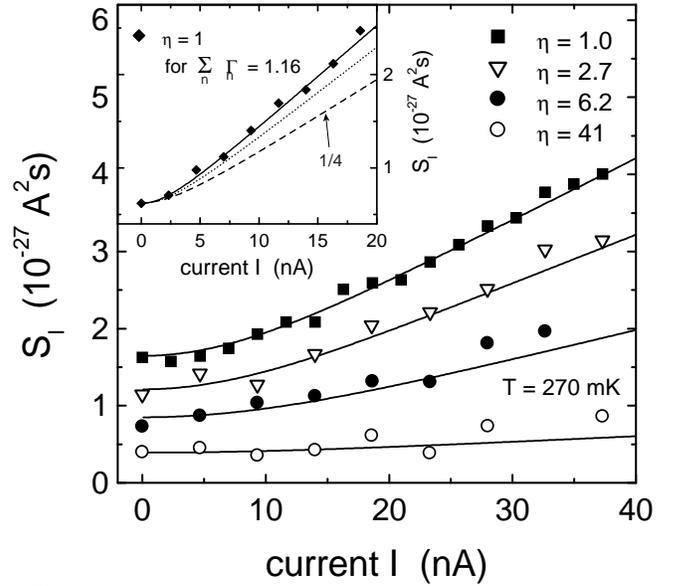}
\vspace{.01mm}
\caption{Shot noise of a chaotic cavity with ideal contacts 
($G_{L,R}/G_{0}=integer$) for different conductance ratios 
$\eta=G_{L}/G_{R}$. The data have an offset for clarity. 
Inset: Shot noise is larger than $1/4\cdot 2e|I|$ if there is 
additional partitioning due to non-ideal contacts ($G_{L,R}/G_{0}\neq 
integer$). The curves are numerical calculations 
assuming no mode mixing (dotted) and for sligth mode mixing of 10\% (solid).}
\label{cavitynoise}
\end{figure}\noindent 
non-equilibrium phenomenon observed in its purest form in the limit 
$eV \gg k_{B}T$, in this experiment bias voltages are limited to $\simeq$8$k_{B}T/e$, only. 
This is to avoid non-linearities of the current-voltage characteristics of the QPCs 
\cite{Kouwenhoven1989} and 1/f-noise-contributions occuring at larger currents \cite{Kumar96}. 
Within this limit, the differential resistance, recorded for all noise measurements, 
changes by $\alt$2.5 $\%$. 
The current noise is finally obtained from the measured voltage 
fluctuations by $S_{I}=S_{V}/(dV/dI)^2-S_{I}^{off}$.

Fig.~\ref{cavitynoise} shows shot noise measurements of a cavity defined by gates 
A and B with a 
size of \mbox{$\simeq$5$\times 8$\,$\mu$m} for different 
symmetry parameters $\eta=G_{L}/G_{R}$.
The solid curves describe the crossover from thermal to 
shot noise for the measured value of $\eta$ given by \cite{BSPRL2000}
\begin{equation}
	S = S_{eq}\left\{1+F(\eta)\cdot\left[\frac{eV}{2k_{B}T}
	\coth\left(\frac{eV}{2k_{B}T}\right)-1\right]\right\}.
	\label{eq:fittingcold}
\end{equation}
$S_{eq}=4k_{B}TG$ denotes the equilibrium noise and $F(\eta)$ the 
Fano factor (Eq.~(\ref{eq:cavitynoisecold})). 
In the symmetric case ($\eta=1$) with $N_{L}=N_{R}=5$
we obtain a very good agreement between the experimental data and the theoretical 
prediction of $1/4 \cdot 2e|I|$. When the right contact is further 
opened ($G_{R}>G_{L}$)  $\eta$ increases
from $1$ (symmetric) to $\simeq$41 (asymmetric). Thereby, shot noise 
gradually disappears for larger values of
$\eta$ as expected from Eq.~(\ref{eq:cavitynoisecold}).
For partial transmission in the contacts shot noise is larger 
than $1/4 \cdot 2e|I|$ because additional noise is generated at the contacts.
This is shown in the inset of Fig.~\ref{cavitynoise} 
where the first mode in the contacts is fully transmitted ($\Gamma_{1}=1$) while 
the second one is partially reflected ($\Gamma_{2}=0.16$). The curves are numerical 
calculations for no mode mixing (dotted) and for slight mode mixing of 
$\simeq 10\%$ (solid) with $\Gamma_{1}=0.90$ and $\Gamma_{2}=0.26$.

\begin{figure}[h]
\centering
\epsfxsize=86 mm
\epsfbox{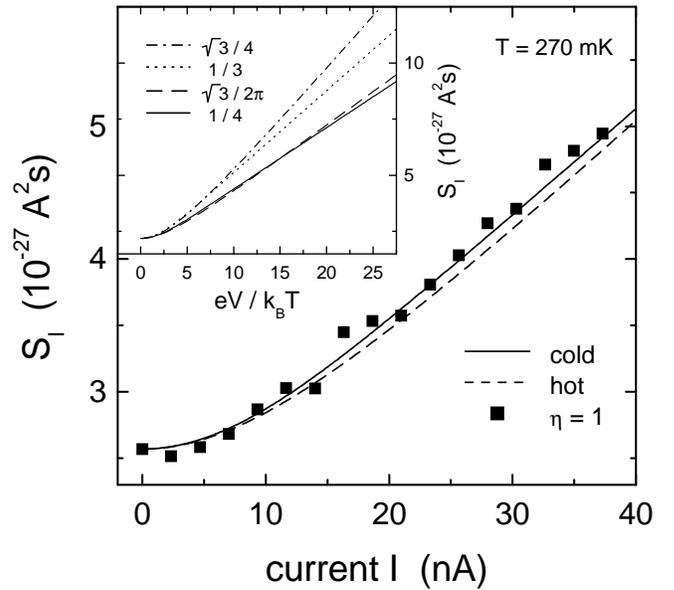}
\vspace{.01mm}
\caption{Shot noise of a symmetric cavity and theoretical predictions 
for cold (solid) and hot electrons (dashed). Inset: comparison of the 
noise of a chaotic cavity ($1/4$ and $\sqrt{3}/2\pi$) with a
diffusive wire ($1/3$ and $\sqrt{3}/4$) for cold and hot electrons.}
\label{coldorhot}
\end{figure}

Up to now we have assumed that inelastic electron scattering inside
the cavity can be neglected. 
In general, heating caused by electron-electron interaction 
enhances shot noise \cite{review}.  
The Fano factor of a diffusive wire, for example, 
changes from $1/3$ for non-interacting (cold) 
electrons to $\sqrt{3}/4$ for 
interacting (hot) electrons \cite{Nagaev95}. 
Heating also affects the shot noise of a chaotic cavity. 
The Fano factor is modified to \cite{deJongReview}:
\begin{equation}
	F(\eta)=\frac{\sqrt{3N_{L}N_{R}}}{\pi(N_{L}+N_{R})}
	=\frac{\sqrt{3\eta}}{\pi(1+\eta)},
	\label{eq:cavitynoisehot}
\end{equation}
and the crossover from thermal to shot noise is described by 
\begin{equation}
		S =\frac{S_{eq}}{2}\left\{1+\sqrt{1+F(\eta)^2\cdot
	\left(\frac{eV}{k_{B}T}\right)^2}\right\}.
	\label{eq:fittinghot}
\end{equation}
For a  symmetric cavity $F(\eta$=$1)\simeq 0.276$ 
for hot electrons, which
is only slighly larger than $F(\eta$=$1)=0.25$ for
cold electrons.
The inset of Fig.~\ref{coldorhot} compares $S(eV/k_{B}T)$
in the hot and cold electron regime for a diffusive
wire and a cavity. As is evident, the differences
are very small, in particular in case of a cavity where
even a crossing at $eV/k_{B}T\simeq 15$ occurs.
In Fig.~\ref{coldorhot} the measured noise 
for $\eta=1$
of Fig.~\ref{cavitynoise} is 
replotted and compared to the 
prediction for cold (solid)
and for hot electrons (dashed). Although the data points
lie clearly closer to the prediction for cold electrons, this
alone is not sufficient to decide which regime is realized in the 
cavity, because of the finite experimental accuracy.
An additional criterion is needed.

\begin{figure}[h]
\centering
\epsfxsize=86 mm
\epsfbox{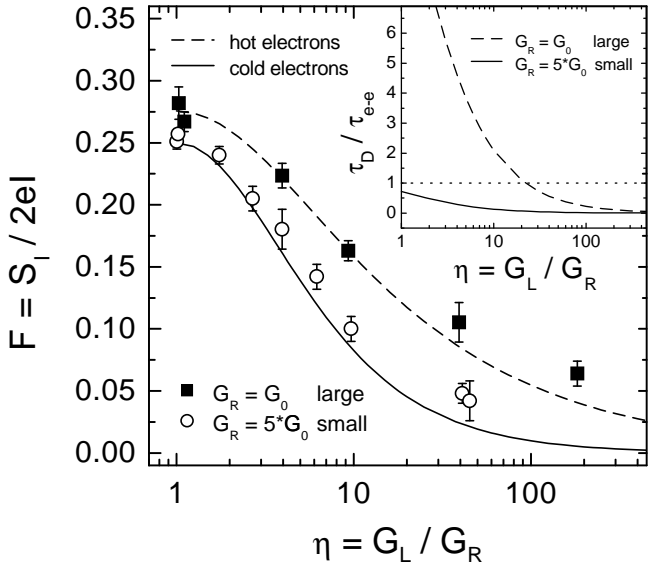}
\vspace{.01mm}
\caption{Fano factor $F\equiv S/2e|I|$ vs. the symmetry parameter $\eta$ for 
(open circles) a small cavity with widely opened contacts ($\tau_{D}<
\tau_{ee}$) and for (black squares) a large cavity with nearly closed contacts 
($\tau_{D}\gg \tau_{ee}$). Predictions for cold electrons (solid) and 
hot electrons (dashed).
Inset: $\tau_{D}/\tau_{ee}$ vs. $\eta$ for the two different types of 
cavity.}
\label{Fanovseta}
\end{figure}

In order to decide whether the cold or hot electron 
theory is appropriate for the comparison with
the measurements, the electron-electron scattering
time $\tau_{ee}$
is compared with the dwell time for electrons inside
the cavity.
We argue that thermalization 
is present 
if $\tau_{D}\gg \tau_{ee}$. 
The average dwell time is the product of the  ballistic flight 
time across the cavity $\tau_{F}\simeq L/v_{F}$ with
the number of scattering events 
inside the cavity given by the ratio
 of the cavity size $L$ and the width of the contacts 
$W=W_{L}+W_{R}=\frac{\lambda_{F}}{2}(N_{L}+N_{R})$:
\begin{equation}
	\tau_{D}=\frac{2\pi\hbar}{E_{F}}
	\left(\frac{L}{\lambda_{F}}\right)^2\frac{1}{(N_{L}+N_{R})}.
	\label{eq:dwelltime}
\end{equation}
The electron-electron scattering rate $\tau_{ee}^{-1}$ in a two 
dimensional electron system is given by \cite{Giuliani1982}
\begin{equation}
	\tau_{ee}^{-1}=\frac{E_{F}}{2\pi\hbar}\left(\frac{k_{B}T_e}{E_{F}}
	\right)^2\left[ \ln\left(\frac{E_{F}}{k_{B}T_e}\right) +  
	\ln\left(\frac{2q}{k_{F}}\right) + 1 \right]
	\label{tee}
\end{equation}
with the Thomas-Fermi screening wave vector 
$q=2me^2/\epsilon_{r}\epsilon_{0}\hbar^2$. 
Because the system is out
of equilibrium the temperature $T_e$
in Eq.~(\ref{tee}) has to be replaced by the effective
electron temperature $T_{eff}$ given by 
\mbox{$T_{eff}=(1/k_{B})\int d\epsilon f_{C}(1-f_{C})$} 
\cite{comment2}.
The ratio $\tau_{D}/\tau_{ee}$ is plotted in the inset 
of Fig.~\ref{Fanovseta} as a function of $\eta=G_{L}/G_{R}$
 \noindent for the
two different types of cavities taking $\tau_{ee}$ from Eq.~(\ref{tee})
for $T_{eff}$ corresponding to the \emph{largest} 
applied voltage $V$ in the experiment.
The upper curve belongs to the large cavity ($\simeq$11$\times$8 $\mu$m), where 
the right contact is nearly closed ($G_{R}$ fixed to $G_{0}$). In this case, 
$\tau_{D} \gg \tau_{ee}$. The lower curve corresponds to the
smaller cavity \mbox{($\simeq$5$\times$8 $\mu$m)} 
with a $5$ times larger opening of the right contact. For this 
type of cavity we find $\tau_{D} < \tau_{ee}$.

According to this argument we use Eq.~(\ref{eq:fittinghot}) valid 
for hot electrons to fit the noise data obtained for 
chaotic cavities with $\tau_{D}/\tau_{ee}>1$. 
The Fano factor $F$ is the only 
fitting parameter. On the other hand, we use
Eq.~(\ref{eq:fittingcold}) valid for cold electrons
if $\tau_{D}/\tau_{ee}<1$.  
The Fano factors  $F=S/2e|I|$ obtained according
to this procedure are plotted as 
a function of the measured $\eta$ for the two 
different cavities described above. 
For the black squares, which belong to the large 
cavity with nearly closed contacts (large dwell time),
we find good agreement with the theoretical Fano factor 
for hot electrons 
given by Eq.~(\ref{eq:cavitynoisehot}) (dashed). 
The open circles are results for the small cavity with wider opened
contacts (small dwell time)
which are consistent with  the prediction for 
non-interacting  electrons described by Eq.~(\ref{eq:cavitynoisecold}).
If we use the formula for cold electrons instead of 
the one for hot-electrons to fit the data obtained for the larger cavity,
the black squares move only slightly downwards by 
$\simeq 0.02-0.03$. They still lie clearly above the 
open circles, demonstrating that heating is indeed
important for the larger cavity.
Good agreement between theory and experiment is found
for both regimes
with the exception of very asymmetric contacts, i.e.
$\eta\gg 1$.
Here, we attribute the deviations to slight mode mixing
within the QPCs, which is difficult to avoid \cite{Kumar96}. 
Let us assume, as an example for the data point at $\eta=180$, 
that two modes instead of one participate in the left
 \noindent
contact 
transmitting respectively with $\Gamma_{1}=0.97$ 
and $\Gamma_{2}=0.03$ instead of $\Gamma_{1}=1.00$ and 
$\Gamma_{2}=0$. This 
yields a Fano factor of $\simeq 0.06$ 
in agreement to what is experimentally observed. 

In conclusion, we have experimentally studied shot noise of
open chaotic cavities defined by two QPCs in series. 
In the regime of 
non-interacting electrons a  Fano factor $F=S/2e|I|$ of 
$1/4$ has been measured as theoretically predicted
for symmetric cavities. 
The origin of this shot noise is
partitioning of the electron wave function by 
quantum-mechanical diffraction inside the 
cavity.  The contacts themselves, which actually 
define the resistance of the 
system, do {\em not} contribute to noise.
In addition, we have also investigated heating effects due to 
inelastic electron-electron scattering by changing the opening of the 
contacts as well as the size of the cavity. 
Similar to other mesoscopic 
systems heating increases shot noise
in agreement with theory.
Shot noise in chaotic cavities is a purely quantum
phenomenon. It would be
interesting to study the crossover from 
``quantum chaos'' to ``classical chaos'', 
where shot noise is predicted to
be absent \cite{Agam}.

The authors would like to thank Ya.~M.~Blanter for valuable comments.
This work was supported by the Swiss National Science Foundation.


\newpage

\begin{thebibliography}{10}
\bibitem{review}
  For a recent review, see: Ya.~M.~Blanter and
  M.~B\"uttiker, "Shot Noise in Mesoscopic Conductors", cond-mat/9910158.
  	
\bibitem{Lesovik1989}
  G.~B.~Lesovik, JETP Lett. 49, 592 (1989).
  
\bibitem{Buttiker1990}
  M.~B\"uttiker, Phys.\ Rev.\ Lett.\ {\bf 65}, 2901 (1990).
  
\bibitem{Jalabert94}
  R.~A.~Jalabert, J.-L.~Pichard, C.~W.~J.~Beenakker, 
  Europhys.\ Lett.\ {\bf 27}, 255 (1994). 
  
\bibitem{BSPRL2000}
  Ya.~M.~Blanter and E.~V.~Sukhorukov,
  Phys.\ Rev.\ Lett.\ {\bf 84}, 1280 (2000).
  
\bibitem{ButtikerBeenakker92}
  C.~W.~J.~Beenakker and M.~B\"uttiker, Phys.\ Rev.\ B {\bf 46}, 1889 (1992);
  K.~E.~Nagaev, Phys.\ Lett.\ A {\bf 169}, 103 (1992).
 
\bibitem{Schoelkopf97}
  R.~J.~Schoelkopf \emph{et al.}, Phys.\ Rev.\ Lett. {\bf 78}, 3370 (1997).  
  
\bibitem{onethird}
  M.~Henny, S.~Oberholzer, C.~Strunk, C.~Sch\"onenberger, 
  Phys.\ Rev.\ B {\bf 59}, 2871 (1999).
  
\bibitem{Nazarov94}
  Yu.~V.~Nazarov,
  Phys.\ Rev.\ Lett.\ {\bf 73}, 134 (1994).
  
\bibitem{SukhorukovPRL}
  E.~V.~Sukhorukov and D.~Loss, Phys.\ Rev.\ Lett. {\bf 80}, 4959 
  (1998).  
  
\bibitem{Wees1988}
  B.~J.~van Wees \emph{et al.}, Phys.\ Rev.\ Lett.\ {\bf 60}, 848 (1988);
  D.~A.~Wharam \emph{et al.}, J.\ Phys.\ C\ {\bf 21}, L209 (1988).
   
\bibitem{Landauer1957}
  R.~Landauer, IBM J.\ Res.\ Dev. {\bf 1}, 223 (1957). 
  
\bibitem{Marcus98}
  A.~G.~Huibers, M.~Switkes, C.~M.~Marcus, K.~Campman, A.~C.~Gossard,
  Phys.\ Rev.\ Lett.\ {\bf 81}, 1 (1998).
  
\bibitem{comment}
  Weak localization can be neglected due to residual magnetic flux 
  through the cavity what is larger than $\phi_{0}$.

\bibitem{Kumar96}
  A.~Kumar, L.~Saminadayar, D.~C.~Glattli, Y.~Jin, and B.~Etienne,
  Phys.\ Rev.\ Lett.\ {\bf 76}, 2778 (1996).
  
\bibitem{Kouwenhoven1989}
  L.~P.~Kouwenhoven \emph{et al.}, Phys.\ Rev.\ B {\bf 39}, 8040 (1989).
   
\bibitem{Nagaev95}
  K.~E.~Nagaev, Phys.\ Rev.\ B {\bf 52}, 4740 (1995);
  V.~I.~Kozub and A.~M.~Rudin, Phys.\ Rev.\ B {\bf 52}, 7853 (1995);
  A.~H.~Steinbach, J.~M.~Martinis, and M.~H.~Devoret,
  Phys.\ Rev.\ Lett.\ {\bf 76}, 2778 (1996).  

\bibitem{deJongReview}
  M.~J.~M.~de Jong and
  C.~W.~J.~Beenakker in {\em Mesoscopic Electron Transport},
  L.~P.~Kouwenhoven, G.~Sch\"on, L.~L.~Sohn eds.,
  NATO ASI Series E, Vol. 345 (Kluwer Academic, Dordrecht 1996).

\bibitem{Giuliani1982}
  G.~F.~Giuliani and J.~J.~Quinn, Phys.\ Rev.\ B {\bf 26}, 4421 (1982).

\bibitem{comment2}
  Thermalization has been assumed for $f_{C}$ which could slightly 
  overestimate $\tau_{ee}^{-1}$.
  
\bibitem{Agam}
  O.~Agam, I.~Aleiner, A.~Larkin, cond-mat/9912086.
\end{thebibliography}
\end{document}